\documentclass{osa-article}

\journal{oe}

\articletype{Research Article}

\usepackage{color}
\usepackage{graphicx}
\usepackage{dcolumn}
\usepackage{bm}

\newcommand{\ket}[1]{| #1 \rangle}

\newcommand{\creatop}[1]{\hat{a}^{\dagger}_{#1}}

\begin{document}

\title{Engineered photon-pair generation by four-wave mixing in asymmetric coupled waveguides.}

\author{Robert J.~A.~Francis-Jones,\authormark{1,2,*} Thomas A.~Wright,\authormark{1} Andriy V.~Gorbach,\authormark{1} and Peter J.~Mosley\authormark{1}}

\address{\authormark{1}Centre for Photonics and Photonic Materials, Department of Physics, University of Bath, Bath, BA2 7AY, UK\\
\authormark{2}Clarendon Laboratory, University of Oxford, Parks Road, Oxford, OX1 3PU, UK}

\email{\authormark{*}jamie.francis-jones@physics.ox.ac.uk} 



\begin{abstract}
Third-order nonlinear processes require phase matching between the interacting fields to achieve high efficiencies. Typically in guided-wave $\chi^{(3)}$ platforms this is achieved by engineering the dispersion of the modes through the transverse profile of the device. However, this limits the flexibility of the phase matching that can be achieved. Instead, we analyze four-wave mixing in a pair of asymmetric waveguides and show that phasematching may be achieved in any $\chi^{(3)}$ waveguide by coupling of a nondegenerate pump from an adjacent waveguide. We demonstrate the additional flexibility that this approach yields in the case of photon-pair generation by spontaneous FWM, where the supermode dispersion may be modified to produce pure heralded single photons -- a critical capability required for example by silicon platforms for chip-scale quantum photonics.
\end{abstract}

\section{Introduction}
Photon-pair generation by parametric nonlinear frequency conversion requires phasematching -- controlling phase velocity so that the sums of the wave-vectors of the pump and photon-pair fields are equal. Early experiments in birefringent nonlinear crystals achieved phasematching by angle tuning \cite{Burnham1970Observation-of-Simultaneity-in-Parametric, Hong1987Measurement-of-subpicosecond-time} but the same technique cannot be used in modern guided-wave nonlinear devices that exploit either $\chi^{(2)}$ or $\chi^{(3)}$ nonlinearity. In second-order nonlinear interactions it is common practice to achieve efficient conversion through quasi-phasematching (QPM) by periodically reversing the sign of $\chi^{(2)}$ \cite{Lim1989Second-harmonic-generation-of-green}. Although the resulting growth is less rapid than that obtained by perfect phasematching, the benefits of confinement to a well-defined set of spatial modes have rendered QPM highly successful in guided-wave nonlinear optics  \cite{URen2004Efficient-Conditional-Preparation, Herrmann2013Post-selection-free-integrated}. Furthermore, QPM is compatible with techniques that limit spectral correlation in photon-pair sources -- crucial for producing high-purity heralded single photons suitable for quantum technologies \cite{Grice2001Eliminating-frequency-and-space-time}. These techniques include both group-velocity matching between the pump and generated fields \cite{Eckstein2011Highly-Efficient-Single-Pass} as well as modulation of the poling period along the length of the waveguide to produce an apodised effective nonlinearity profile \cite{Branczyk2011Engineered-optical-nonlinearity, Graffitti2018Independent-high-purity-photons}.

In contrast to second-order nonlinearity, the symmetry of third-order nonlinearity means that four-wave mixing (FWM) can take place in materials from which waveguides are commonly fabricated, for example silica optical fibre \cite{Li2004All-fiber-photon-pair-source} and silicon photonic chips \cite{Silverstone2014On-chip-quantum-interference}. However, the invariance of $\chi^{(3)}$ under a sign change neutralises the capability of domain reversal to influence third-order nonlinear interactions; hence it cannot be used for QPM. The phasematching conditions in waveguided $\chi^{(3)}$ devices are therefore determined by the material and geometric contributions to the overall dispersion. Hence it is typically the waveguide dispersion that must be modified to achieve phasematching \cite{Sharping2006Generation-of-correlated-photons}.

In some situations, such as FWM in photonic crystal fibre (PCF), the flexibility afforded by engineering the dispersion of a single waveguide is sufficient to access a wide range of target wavelengths that may be far from the pump. For example, in  the case of photon-pair generation with a Nd or Yb pump laser at 1064\,nm one can produce signal photons at 800\,nm which can be detected with high-efficiency silicon avalanche photodiodes with corresponding idler photons suitable for long-distance transmission over fibre in the telecommunication C-band around 1550\,nm \cite{McMillan2009Narrowband-high-fidelity-all-fibre}. Similar wavelength shifts are possible using birefringence in UV-written waveguides in silica \cite{Posner2018High-birefringence-direct-UV-written}. Furthermore, these systems give sufficient flexibility to allow the group velocities of the interacting fields to be manipulated, producing high-purity heralded single photons \cite{Clark2011Intrinsically-narrowband-pair, Francis-Jones2016All-fiber-multiplexed-source, Spring2017Chip-based-array-of-near-identical}.

However in other cases, in particular integrated waveguides fabricated in silicon, there is comparatively little control that can be brought to bear on the properties of the photon pairs produced \cite{Sharping2006Generation-of-correlated-photons}. Although periodic width modulation \cite{Driscoll2012Width-modulation-of-Si-photonic-wires, Bahar2018Adiabatic-four-wave-mixing}, changes of direction in crystalline materials \cite{Driscoll2011Directionally-anisotropic-Si-nanowires:}, and pairs of identical waveguides \cite{Biaggio2014Coupling-length-phase-matching} have been suggested for phase matching $\chi^\text{(3)}$ difference-frequency generation, photon pairs are typically generated close to the pump in the anomalous dispersion regime, where the only detuning is provided by power-dependent cross-phase modulation of the generated fields by the pump. Furthermore, the pairs are typically highly anti-correlated in frequency, degrading the purity of the resulting heralded single photons \cite{Jizan2015Bi-photon-spectral-correlation}. Micro-resonators \cite{Vernon2017Truly-unentangled-photon} and periodically-tapered waveguides \cite{Saleh2018Quasi-phase-matched-chi3-parametric-interactions} have been proposed to yield greater control over the properties of photon pairs, but the resulting devices are complex to fabricate and likely to suffer from additional loss over their uniform counterparts. Symmetric directional couplers have also been studied as a method of generating entangled pairs of photons, though their spectral properties were not controlled \cite{Frank2015Tunable-generation-of-entangled}.

We propose a simple technique that offers more flexible FWM phasematching across a variety of guided-wave platforms while also offering the possibility of control over the nonlinearity profile along the device. Rather than carrying out photon-pair generation by FWM in a single waveguide, we introduce a second waveguide strongly coupled at one of two non-degenerate pump wavelengths. By tuning the amount of coupling between the waveguides we can directly control the supermode dispersion of the  long-wavelength pump to achieve phasematching. Introducing asymmetry between the waveguides enables us to ensure that the coupling is strong only for the long-wavelength pump while the short-wavelength pump remains localised to a single waveguide. Hence, because photon-pairs are generated in the region of overlap between the two pumps, the asymmetry forces the generated fields to remain largely in the fundamental spatial mode of a single waveguide facilitating efficient collection.  We present firstly the analysis of the general case of coupling-induced FWM, and proceed to give details of a specific design to generate photon pairs by FWM in silicon-on-insulator (SOI) waveguides. In this system we show how the supermode dispersion can be engineered to generate photon-pairs that are far detuned from the pumps and also group-velocity matched so that heralded single photons may be produced directly in pure states. Furthermore, we demonstrate that this technique  maintains the exponential growth rate associated with perfect phase matching, rather than the diminished growth seen in QPM-type schemes. Finally, we discuss how our ideas could be extended to produce intensity modulation of the pump, and hence the nonlinearity, along the waveguide, thereby enhancing the purity of heralded single photons even further.

\section{General case}

We consider continuous-wave FWM in a pair of $\chi^{(3)}$ waveguides labelled A and B, driven by two nondegenerate pump fields at wavelengths $\lambda_{p1}$ and $\lambda_{p2}$ where $\lambda_{p1} < \lambda_{p2}$. When the two waveguides are coupled we obtain two possible modes of the system at any given frequency $\omega_j$. These are known as the even and odd supermodes and have propagation constants $\beta_j^{(+)}$ and $\beta_j^{(-)}$ respectively: 
\begin{align}
	\beta_j^{+} &= k_j + \kappa_j,\\
    \beta_j^{-} &= k_j - \kappa_j,
\end{align}
where $k_j = k_{A,j} = k_{B,j}$ are the modes of the isolated waveguides. $\kappa_j$ is the coupling constant at $\omega_j$ and sets the shift in the propagation constant in one waveguide due to the presence of the other. $\kappa_j$ is strongly frequency-dependent, hence it can be seen that by controlling the coupling strength, for example by altering the separation of the waveguides, we can directly affect the dispersion of the supermodes, and phasematch FWM at particular wavelengths of interest.

The spatial distribution of the electric field for the corresponding eigenmodes can be approximated by:
\begin{eqnarray}
	E_{+}(\omega,x,y) \approx \frac{1}{\sqrt{2}}\left[E_{A}(\omega,x,y) + E_{B}(\omega,x,y)\right], \\
	E_{-}(\omega,x,y) \approx \frac{1}{\sqrt{2}}\left[E_{A}(\omega,x,y) - E_{B}(\omega,x,y)\right].
\end{eqnarray}
Where $E_{A}(\omega,x,y)$ and $E_{B}(\omega,x,y)$ is the distribution of the electric field in waveguide A and B respectively. Depending on the launch conditions it possible to excite either the even or odd supermode or any linear combination of the two. If a single supermode is excited, it will propagate through the structure with no change in intensity. However, if a linear combination of supermodes are excited, then they will beat against each other giving a sinusoidally varying field amplitude with a periodicity of $2\pi/\kappa$. If for a given wavelength the coupling is very weak then the beat length can be much larger than the length of the coupled region. If in this case, light is injected into only one of the waveguides, a negligible amount of power will be transferred in to the other waveguide.  

\begin{figure}
\centering\includegraphics[width = 0.65\textwidth]{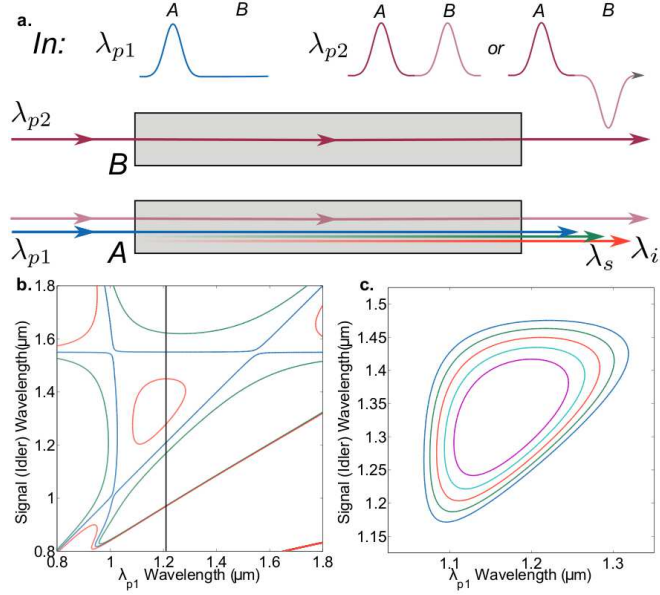}
\caption{a.) Schematic of coupling-induced FWM in a pair of waveguides. A short-wavelength pump at $\lambda_{p1}$ is injected solely into waveguide A in which FWM takes place with a long-wavelength pump at $\lambda_{p2}$ described by even and odd supermodes illustrated above. b.) Phase matching for coupling-induced FWM in a pair of coupled fibre cores with parameters similar to SMF28 telecoms fibre. Contours of zero phase mismatch plotted as a function of $\lambda_{p1}$ with constant $\lambda_{p2} = 1550\,\text{nm}$ for three cases: $\Delta k = 0$ (blue, phase matching only occurs at the pump wavelengths), $\Delta k^{(+)} = 0$ (red), and $\Delta k ^{(-)} = 0$ (green). c.) Series of phasematching contours for increasing value of $\kappa_{p2}$ for the $\Delta k^{(+)} = 0$ case. Increasing $\kappa_{p2}$ causes the contours to collapse towards a single point at which degenerate pairs are produced.}
\label{fig:schematic}
\end{figure}

Initially, we look for phasematching that will enable the spontaneous generation of signal and idler fields, $\lambda_{s}$ and $\lambda_{i}$, where $\lambda_{p1} < \lambda_{s} \leq \lambda_{i} < \lambda_{p2}$ under the conditions that the fields are monochromatic. For each waveguide in isolation, we have the standard condition that for efficient FWM the phase mismatch $\Delta k$ must satisfy:
\begin{equation}
\Delta k = k_{p1} + k_{p2} - k_{s} - k_{i} - \gamma P = 0,
\label{eq:dk}
\end{equation}
and we set the peak power $P$ to be sufficiently low that the term arising from the nonlinearity $\gamma$ can be neglected. We then consider two coupled waveguides sufficiently detuned from one another that we can ignore the coupling for all wavelengths other than the long-wavelength pump $\lambda_{p2}$ (the validity of this assumption will be discussed in more detail later). A schematic is shown in Fig.\,\ref{fig:schematic}(a). In this case there are two possibilities to achieve phase matching:
\begin{align}
	\Delta k^{(+)} &= k_{p1} + k_{p2} + \kappa_{p2} - k_{s} - k_{i} = 0,\label{eq:dk_pos}\\
    \Delta k^{(-)} &= k_{p1} + k_{p2} - \kappa_{p2} - k_{s} - k_{i} = 0.\label{eq:dk_neg}
\end{align}
These correspond to phasematching with the long-wavelength pump in either the even or odd supermode of the coupled system, with $\Delta k^{(+)} = \Delta k + \kappa_{p2}$ and $\Delta k^{(-)} = \Delta k - \kappa_{p2}$.

Fig.\,\ref{fig:schematic}(b) shows an example of the phase matching that results if the two waveguides are a pair of coupled fibre cores with the same parameters as SMF28 telecoms single-mode step-index fibre. We have plotted the loci of points for the three cases: $\Delta k^{(+)}=0$ (red), $\Delta k^{(-)}=0$ (green), and $\Delta k=0$ (blue). In the absence of coupling we see that phase matching only occurs at the two pump wavelengths (recall that we have set $\gamma P \approx 0$). However, by including a small amount of coupling, in this case $\kappa_{p2} = 250 \text{m}^{-1}$ for the two identical fibre cores, we see that two additional solutions arise at each pump wavelength, corresponding to equations \ref{eq:dk_pos} and \ref{eq:dk_neg}. In addition, it is clear that by tuning the pump wavelength $\lambda_{p1}$ the gradient of the phase matching contours can be adjusted. For both $\Delta k^{(+)}=0$ and $\Delta k^{(-)}=0$ the gradient of one contour in Fig.\,\ref{fig:schematic}(b) is zero for a particular value of pump wavelength. At these points, the conjugate photon is group-velocity matched to the pump field; this is one condition required to create photon pairs in factorable states and hence high-purity heralded single photons. Fig.\,\ref{fig:schematic}(c) illustrates a series of $\Delta k^{(+)}=0$ contours as $\kappa_{p2}$ is increased. It can be seen that $\kappa_{p2}$ yields control over the curvature of the contour which can then be exploited for engineering the joint spectral distribution of the photon-pairs. This is discussed in greater detail in Section\,\ref{sec:state_engineering}. Eventually as $\kappa_{p2}$ is increased further, $\Delta k^{(+)}=0$ collapses to a single point, at which degenerate photon pairs can be produced.

\section{Pulse propagation model}
\label{sec:PulseProp}

We have modelled coupling-induced FWM in our system through a complete numerical solution to the generalised nonlinear Schr\"odinger equation including the coupling of the long wavelength pump to the bus waveguide B. The amplitude of each field in the FWM waveguide was governed by one of a series of coupled differential equations of the form:
\begin{equation}
\frac{dA_{F}}{dz} = i\beta_{F}A_{F}+ i\kappa_{F}A_{F'} + i\gamma \Big( \Psi_F + \Phi_F \Big),
\end{equation}
where the subscript F denotes field P1, P2, S or I, and F' denotes the corresponding components of each field in the adjacent waveguide. The relations for phase modulation, $\Psi_F$, and FWM, $\Phi_F$, are given by:
\begin{equation}
\Psi_F = \Big( |A_F|^2 + 2\sum_{J\neq F}|A_J|^2 \Big)A_{F},
\end{equation}
and
\begin{equation}
 \Phi_F=
    \begin{cases}
      2A_{P2}^*A_S A_I \,e^{i\Delta kz}, & F=P1 \\
      2A_{P1}^*A_S A_I \,e^{i\Delta kz}, & F=P2 \\
      2A_{I}^*A_{P1} A_{P2} \,e^{-i\Delta kz}, & F=S \\
      2A_{S}^*A_{P1} A_{P2} \,e^{-i\Delta kz}, & F=I
    \end{cases}
\end{equation}
respectively. The equations were solved in the retarded frame of pump 2, with the group velocity of this field subtracted from the propagation of all pulses. Furthermore, the coefficients $\kappa_{P1,S,I}$ were assumed to be sufficiently small that coupling between the waveguides could be neglected at these wavelengths.

We use the split-step Fourier method to propagate the broadband short-wavelength pump pulses launched into the fundamental mode of FWM waveguide (A) and CW long-wavelength pump light in any linear combination of the even and odd supermodes of both waveguides to study the effect this has on the growth of signal and idler fields.

\begin{figure}
\centering\includegraphics[width = 0.65\textwidth]{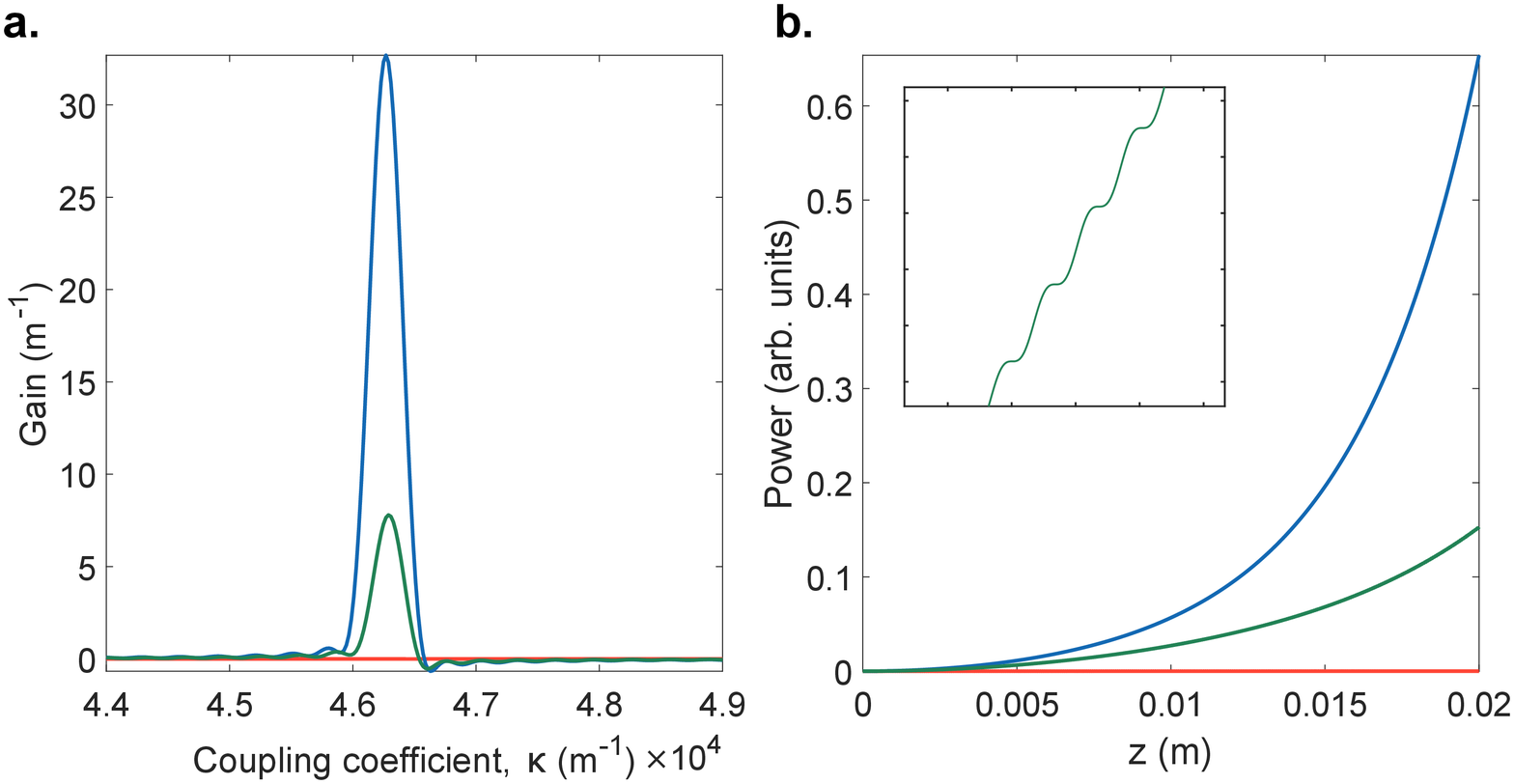}
\caption{a.) FWM gain spectrum as a function of the coupling coefficient $\kappa_{p2}$, for three different launch conditions, even supermode (red), odd supermode (blue), and beating between odd an even (green). High gain is only achieved over a narrow bandwidth where the phase mismatch between the fields is compensated by the shift in propagation constants due to coupling between the waveguides at $\lambda_{P2}$. b.) Growth of the idler field as a function of length along the fibre; inset shows magnification of the growth for the case where the two supermodes beat against one another. Exponential growth is observed for perfect phasematching when the pump is launched in the odd supermode.}
\label{fig:SSFM_results}
\end{figure}

We show an example case consisting of two identical telecom single-mode fibre cores pumped at $\lambda_{p1} = 532$\,nm and $\lambda_{p2} = 1550$\,nm producing degenerate pairs at $\lambda_{s,i} = 792$\,nm when $\kappa_{p2} \approx 46 \times 10^{3} \text{m}^{-1}$. The coupling range over which amplification takes place was found by scanning the value of $\kappa_{p2}$, and is shown in Figure \ref{fig:SSFM_results}a. Hence we see that power transfer from the pumps to the signal and idler takes place over a range of values of $\kappa_{p2}$ for which the coupling can compensate for a phase mismatch that would not otherwise allow efficient FWM to occur. When the even supermode is excited at the input of the system we observe no FWM growth as we are looking away from the phase matched FWM wavelengths (see Fig.\ref{fig:schematic}). On the other hand, when the odd supermode is excited at the input we observe exponential growth of the signal and idler. Finally, when an equal combination of the supermodes are excited at the input, we observe growth but at a reduced rate as only half the pump field has the correct propagation constant to achieve the target phase matching. In this case the two supermodes beat against one another giving a sinusoidal power transfer between the waveguides, leading to a staircase-like modulation to the FWM growth reminiscent of QPM in $\chi^{(2)}$ materials.

\section{Photon-pair state engineering}
\label{sec:state_engineering}

We now investigate in more detail how the flexibility of FWM in coupled waveguides may be exploited to engineer the two-photon state from spontaneous FWM. It has been shown that asymmetric waveguides may be used to create broadband directional couplers \cite{Lu2015Broadband-silicon-photonic}, but here we focus on pairs of asymmetric waveguides to give a greater level of control over FWM phase matching. Identical waveguides will have a coupling strength that increases monotonically with wavelength (as long as the supermode remains well-localised to the region of the waveguides), however with two different waveguides, we can choose a particular wavelength that is phase matched and therefore strongly coupled while both shorter and longer wavelength experience much weaker coupling. We show how this can be used to engineer the joint spectrum of photon pairs.

Beginning from the interaction Hamiltonian, $H_\text{I} = \epsilon_0 \chi^{(3)} E_\text{p1}^\text{(-)} E_\text{p2}^\text{(-)} E_\text{s}^\text{(+)} E_\text{i}^\text{(+)}$, we find the two-photon component of the state generated by FWM is given by:
\begin{equation}
\ket{\psi} = \iint d\omega_s d\omega_i f(\omega_s, \omega_i) \creatop{s}(\omega_s) \creatop{i}(\omega_i) \ket{0}.
\end{equation}
The joint spectral amplitude $f(\omega_s, \omega_i)$ is formed from the product of the pump function:
\begin{equation}
\alpha(\omega_{s} + \omega_{i}) = \int d\omega'\alpha_{p1}(\omega')\alpha_{p2}(\omega_{s} + \omega_{i} - \omega'), 
\end{equation}
and the phase matching function:
\begin{align}
\phi(\omega_s, \omega_i) = & \frac{1}{2} \int_0^L e^{i(\Delta k + \kappa_{p2})z} + e^{i(\Delta k - \kappa_{p2})z} \\
					\begin{split} = & L\, e^{i(\Delta k + \kappa_{p2})L/2} \text{sinc}\left(\frac{(\Delta k + \kappa_{p2})L}																							{2}\right)\\
					& + L\,e^{i(\Delta k - \kappa_{p2})L/2} \text{sinc}\left(\frac{(\Delta k - \kappa_{p2})L}{2}\right).
                    \end{split}
\end{align}
Here we see that $\kappa_{p2}$ enables us to compensate a non-zero phase mismatch $\Delta k$, and that when the coupling is reduced to zero we recover the standard phase matching function. Furthermore, at any given pair of signal and idler wavelengths only one of the two possibilities, $\Delta k^{(+)}=0$ or $\Delta k^{(-)}=0$, will be satisfied and the other can be ignored.

\begin{figure}
\centering\includegraphics[width = 0.35\textwidth]{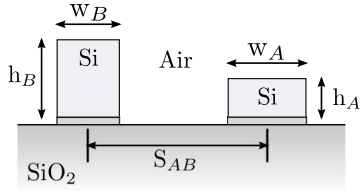}
\caption{Schematic of the coupled asymmetric waveguides. The waveguides are formed of silicon on a silica substrate and sit on a silica pedestal 20nm high.}
\label{fig:asym_guides}
\end{figure}

In addition to simple phase matching requirements, to produce factorable two photon states we also need control over the relative group velocities of the signal, idler, and pump fields. In our system of coupled waveguides, this can be achieved not only by adjusting the parameters of the individual waveguides to control the dispersion, but also through the dispersion in the coupling strength, $\kappa(\omega)$. In this section we show that an asymmetric system gives the flexibility required to achieve both phase- and group-index matching of the fields by varying the cross-section of each waveguide independently; this situation is illustrated in Fig.\,\ref{fig:asym_guides}. In order to model this accurately, we must consider the supermode dispersion of all four fields.

There is an additional advantage of the asymmetric system. Although the coupling at $\lambda_{p2}$ must be strong to give sufficient influence over the supermode dispersion, in order to generate photon pairs in useful spatial modes it is important to ensure that the signal and idler are predominantly localised in one waveguide at the end of the device. If the detuning of $\lambda_{s}$ and $\lambda_{i}$ from $\lambda_{p2}$ is large, this condition is satisfied straightforwardly as the signal and idler will remain in the same waveguide as  the short-wavelength pump $\lambda_{p1}$. However, in the specific example contained here of pair-generation in silicon, all wavelengths must be greater than 1\,$\mu$m yet we wish to generate pairs at telecoms wavelengths to access efficient single-photon detectors. Hence the detuning of the idler from $\lambda_{p2}$ is not large and limiting the coupling at $\lambda_{i}$ presents a challenge. As we will show, this difficulty is mitigated by working with asymmetric waveguides. 

Returning to coupled mode theory, we now consider a system of asymmetric waveguides in which the propagation constants of the even and odd supermodes at frequency $\omega_j$ are given by:
\begin{align}
\beta^{+}_j &= \bar{\beta_j} + \psi_j, \label{eq:smode_1}\\
\beta^{-}_j &= \bar{\beta_j} - \psi_j, \label{eq:smode_2}
\end{align}
where $\bar{\beta_j} = (\beta_{A,j} + \beta_{B,j})/2$ is the mean propagation constant, $\psi_j = (\delta\beta_j^{2} + \kappa_j^{2})^{1/2}$, $\delta\beta_j = (\beta_{A,j} - \beta_{B,j})/2$, and the quantity $\psi_j = (\beta_j^{+} - \beta_j^{-})/2$ can be determined by the difference between the supermode propagation constants\,\cite{chaung1995Physics_of_Optoelectronic_devices}. The spatial distribution of the electric field of the corresponding eigenmodes is approximated by:
\begin{eqnarray}
	E_{j}^{(+)} \approx \frac{1}{N_{j}}\left(\frac{\delta\beta_{j} + \psi_{j}}{-\kappa_{j}}E_{A,j} + E_{B,j}\right),\\
	E_{j}^{(-)} \approx \frac{1}{N_{j}}\left(\frac{\delta\beta_{j} - \psi_{j}}{-\kappa_{j}}E_{A,j} + E_{B,j}\right),
\end{eqnarray}
where $N_{j} = \sqrt{((\delta \beta_{j} \pm \psi_{j})/-\kappa_{j})^{2} + 1^{2}}$ is a normalisation factor. These expressions can be used to design devices that satisfy phase matching for particular sets of pump, signal, and idler wavelengths. 

In general, energy conservation leads to strong anti-correlation between the frequencies of signal and idler, which is undesirable for heralded single-photon generation due to the commensurate requirement for tight spectral filtering to obtain high-purity heralded single photons. It has been demonstrated that two-photon states that are largely devoid of frequency correlation can be generated directly by satisfying constraints on the group velocities of the fields such that, in the degenerate-pump case, $v_{g,s} \leq v_{g,p} \leq v_{g,i}$ and pumping with broadband pulses.

\begin{figure}
\centering\includegraphics[width = 0.65\textwidth]{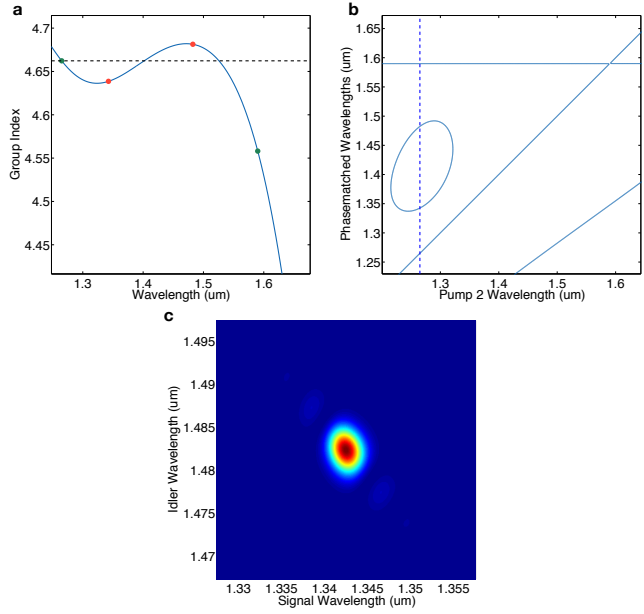}
\caption{a.) Group index profile of even supermode, green circles and dashed line refer to the group index at the pump wavelengths, red circles depict the group index of the daughter photons. Just as with PDC for GVM states the group index of the pump must bisect the group index of the daughter pair. b.) Phasematching contours for the even supermode dispersion, dashed line indicates the the wavelength of the short pump c.) Joint spectral intensity distribution of the generated pairs for $\sigma_{p1} = 2$\,nm and $L = 15$\,mm.}
\label{fig:gvm}
\end{figure}

Satisfying the conditions for high-purity single-photon generation places constraints on material use and device design, as the degrees of freedom available for engineering group velocity in a single waveguide are limited. In particular, for pair-generation in silicon waveguides it is typically not possible to achieve the requisite group-velocity matching required. However, the additional flexibility provided by our technique of coupling-induced FWM, which unlocks phase matching across a broader range of parameters, can enable group-velocity matching to be satisfied even in silicon waveguides. Nevertheless, to produce a spectrally factorable state, we must allow at least one pump field to be broadband. In order that the dispersion of coupling does not affect the long-wavelength pump, we elect to leave the long-wavelength pump monochromatic and use a pulsed short-wavelength pump. In this case, the requirement on the group velocities becomes $v_{g,s} \leq v_{g,p1} \leq v_{g,i}$.We demonstrate one such case in Figure \ref{fig:gvm},  for a system with parameters as detailed in Table\,\ref{tab:Asym_Params} pumped at $\lambda_{p1} = 1.265 \mu$m and $\lambda_{p2} = 1.590 \mu$m.

\begin{table}
\caption{System Parameters for asymmetric coupled waveguides. All dimensions and wavelengths given in $\mu$m.}
\label{tab:Asym_Params}
\centering
\vspace{1.5mm}
\begin{tabular}{|c|c|c|c|c|}
\hline
$w_{A}$ & $h_{A}$ & $w_{B}$ & $h_{B}$ & $S_{AB}$\\
\hline
0.32 & 0.22 & 0.4 & 0.42 & 0.6 \\
\hline
\end{tabular}

\begin{tabular}{|c|c|c|c|}
\hline
$\lambda_{p1}$ & $\lambda_{p2}$ & $\lambda_{s}$ & $\lambda_{i}$ \\
\hline
$1.265$ & $1.590$ & $1.342$ & $1.482$ \\
\hline
\end{tabular}
\end{table}

Fig.\,\ref{fig:gvm}(a) shows the group index for the odd supermode of the structure as a function of wavelength. The points highlighted in red indicate the group index of the two pump fields, and the points in green indicate the group index of the signal and idler photons. In order for the process to result in a factorable state, the group index of one of the pumps must fall between the group indices of the signal and idler; this is reminiscent of the condition for factorable photon-pair generation in parametric downconversion. We see that coupling-induced FWM enables GVM to be satisfied due to the presence of a local minimum in the group index near the signal wavelength followed by a local maximum near the idler. Fig.\,\ref{fig:gvm}(b) illustrates the resulting joint spectral intensity with a pump bandwidth of $\sigma_{p} = 2$nm and interaction length $L = 15$mm. The tight confinement of the short-wavelength pump means that the region of overlap between both pumps is in one waveguide only; hence the photon-pairs are generated in a single waveguide. The asymmetry between the waveguides limits the coupling of both signal and idler, and hence the photon pairs are delivered in a mode that is suitable for coupling to a single-mode waveguide or fibre.

\section{Further discussion}

While fulfilling the group-velocity matching condition eliminates the majority of frequency correlation from the two-photon state, the purity of the heralded single photons still remains limited by residual correlation due to the wings of the sinc phase matching function as seen in Fig.\,\ref{fig:gvm}. These appear due to the sharp boundaries of the interaction region in the direction of propagation; if the boundaries can be blurred the side lobes are suppressed, in analogy with the apodization of a telescope aperture. There exist a number of proposals and implementations of such techniques, for example by modulating QPM periods or gradually switching off nonlinearity by ion implantation in $\chi^{(2)}$ materials, however no techniques exist to perform this apodization in integrated $\chi^{(3)}$ devices.

The additional flexibility afforded by our bus waveguide enables the effective nonlinearity to be switched on and off smoothly by varying the depth of intensity modulation along the length of waveguide A. This could be achieved by detuning the coupling between the two waveguides, by increasing $\delta \beta$ or by reducing $\kappa$. This would amount to increasing the centre-to-centre separation of the guides or changing the width of the bus waveguide, while keeping the FWM phasemismatch $\Delta \beta$ must remain relatively constant.
\begin{figure}
\centering
\includegraphics[width = 0.45\textwidth]{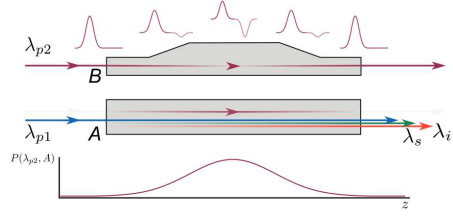}
\caption{Illustration demonstrating how apodisation of the effective nonlinearity may be achieved by slowly modulating the width of one waveguide.}
\label{fig:AppoSchematic}
\end{figure}

So far, we have only considered the use of silicon waveguides on a silica substrate where the ``cladding" on the remaining three surfaces is air. By including a capping layer of silica further control over the dispersion of the system could be achieved, and the same concepts could be applied to other $\chi^\text{(3)}$ materials such as silicon nitride. Additionally, we have presented only one particular set of pump wavelengths and waveguide structural parameters; a broader study into different combinations of waveguide widths, heights, and separations together with appropriate pump wavelengths could lead to additional device functionality. Similar studies of the properties of couplers have made use of genetic algorithms in order to optimise the device design \cite{Fu2016Broadband_optical_waveguide}; similar techniques could be applied here.

\section{Conclusion}
In conclusion, we have presented a novel technique for phase matching nonlinear processes in optical waveguides. By introducing a second bus waveguide that is coupled to the generating guide we have shown that it is possible to have sufficient control over the supermode dispersion to target and phase match these processes at particular wavelengths. We have applied this technique to four wave mixing with a particular emphasis on the process of spontaneous generation of high purity heralded single photons at telecom wavelengths in silicon waveguides. By using the extra degree of flexibility afforded by the introduction of the second waveguide we have been able to simultaneously phasematch FWM but also group-velocity match the pump, signal, and idler. We have also briefly outlined a possible route by which the remaining spectral correlations between signal and idler photons could be removed by apodising the nonlinearity through the structure of the coupled system, which is not a possibility in current silicon quantum photonics platforms.

\section*{Acknowledgments}

This work was funded by the UK EPSRC Quantum Technology Hub Networked Quantum Information Technologies, Grant No. EP/M013243/1.

\bibliography{fwm_couplers}

\end{document}